\documentclass[structabstract, letter]{aa} 
\usepackage{graphicx} 
\usepackage{subfigure}
\usepackage{float}
\usepackage{txfonts}
\usepackage{natbib}
\usepackage{units}
\usepackage{url}
\newcommand{\lsp}{LS~I~+61$^{\circ}$303}
\newcommand{\lsi}{LS~I~+61$^{\circ}$303~}

\newcommand{\beq}{\begin{equation}}
\newcommand{\eneq}{\end{equation}}
\begin{document}

\title{Long-term  OVRO monitoring of \lsp: confirmation of the two close periodicities} 
\author{
        M.\ Massi\inst{1},
        F.\ Jaron\inst{1},
        T.  Hovatta\inst{2, 3}
}

\institute{
  Max-Planck-Institut f\"ur Radioastronomie, Auf dem H\"ugel 69,
  D-53121 Bonn, Germany \\
  \email{mmassi,  fjaron,  @mpifr-bonn.mpg.de}
    \and
  Aalto University 
          Mets\"ahovi Radio Observatory, Mets\"ahovintie 114,
          FIN-02540 Kylm\"al\"a, Finland \\
  \email{talvikki.hovatta@aalto.fi}
  \and
 Cahill Center for Astronomy and Astrophysics, California Institute of 
 Technology, Pasadena CA,  91125, USA\\
 }

\date{Received  }

\abstract
{ 
        The gamma-ray binary  \lsi shows multiple periodicities.
  The timing analysis of 6.7 yr of GBI radio data 
and of 6 yr of
\textit{Fermi}-LAT GeV gamma-ray data  both have found two close periodicities
$P_{1, {\rm GBI}} = 26.49 \pm 0.07$ d, 
       $P_{2, {\rm GBI}} = 26.92 \pm 0.07$ d
and  $P_{1, \gamma} = 26.48 \pm 0.08$ d, $P_{2, \gamma} =
  26.99 \pm 0.08$  d.
}
{ 
The system \lsi is the object of several continuous monitoring programs at low and high energies. 
  The frequency difference between  $\nu_1$ and $\nu_2$ of only
  0.0006~d$^{-1}$ requires long-term monitoring because  
  the frequency resolution in timing analysis  is related to the inverse of the overall time interval. 
The  Owens
Valley Radio Observatory (OVRO) 40~m telescope has been monitoring the source at 15~GHz for 
 five~years and  overlaps with \textit{Fermi}-LAT monitoring.
  The aim of this work is to establish whether the two frequencies are also 
  resolved in the OVRO monitoring.
}
{ 
        We analysed OVRO data with the Lomb-Scargle method. 
        We also updated the timing analysis  of \textit{Fermi}-LAT observations.
}
{ 
        The periodograms of OVRO data confirm the two periodicities $P_{1, {\rm OVRO}} = \unit[26.5 \pm
        0.1]{d}$ and $P_{2, {\rm OVRO}}  = \unit[26.9 \pm 0.1]{d}$. 
}
{ 
        The three indipendent measurements of $P_1$ and $P_2$ with GBI,
        OVRO, and \textit{Fermi}-LAT observations
confirm  that the periodicities  are  permanent features of the system \lsp.
The similar behaviours of the emission  at high (GeV) and low (radio) energy   
when the compact object in \lsi{} is toward apastron
suggest  that  the emission  is  caused by the   same periodically ($P_1$) 
ejected population of electrons in a precessing ($P_2$) jet.
} 
\keywords{Radio continuum: stars - X-rays: binaries - X-rays:
  individual (\lsi) - Gamma-rays: stars}
\titlerunning{OVRO monitoring of  \lsp}
\maketitle
\section{Introduction}

The stellar system \lsi{} is formed by a compact object and a Be star
in an eccentric orbit \citep[$e = 0.72 \pm 0.15$,
][]{casares05}. 
          \textit{Fermi}-LAT observations \citep{abdo09} reveal a large
outburst toward  periastron \citep[$\Phi_{\rm periastron} = 0.23 \pm 0.02$,][]{casares05}
along with  a smaller outburst toward apastron
\citep{jaronmassi14}. Radio observations of the system show a large
radio outburst toward apastron 
exhibiting   a   modulation of $\sim 1667$ days
\citep[][and here Fig.1]{paredes90, gregory02, massikaufman09}. 
\begin{figure*}[ht]
  \centering
  \includegraphics[scale=.45, clip]{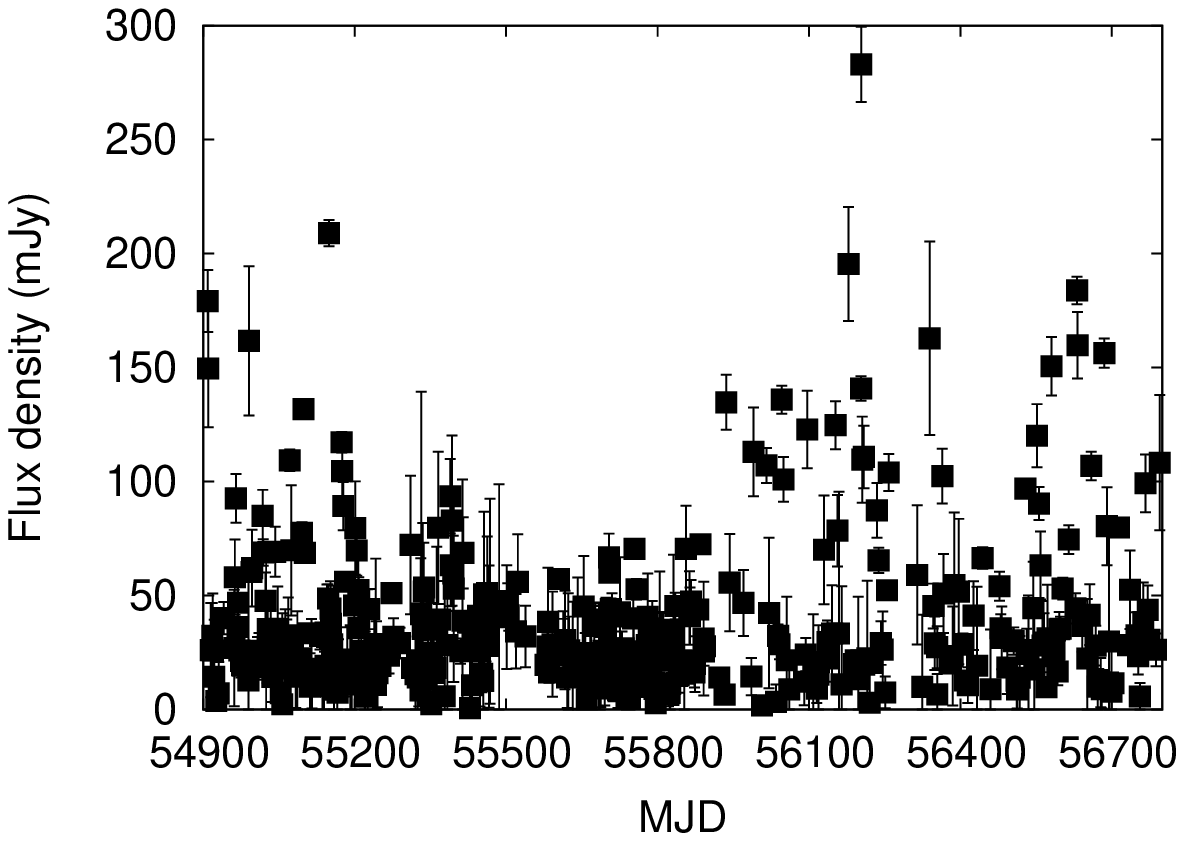}
  \includegraphics[scale=.45, clip]{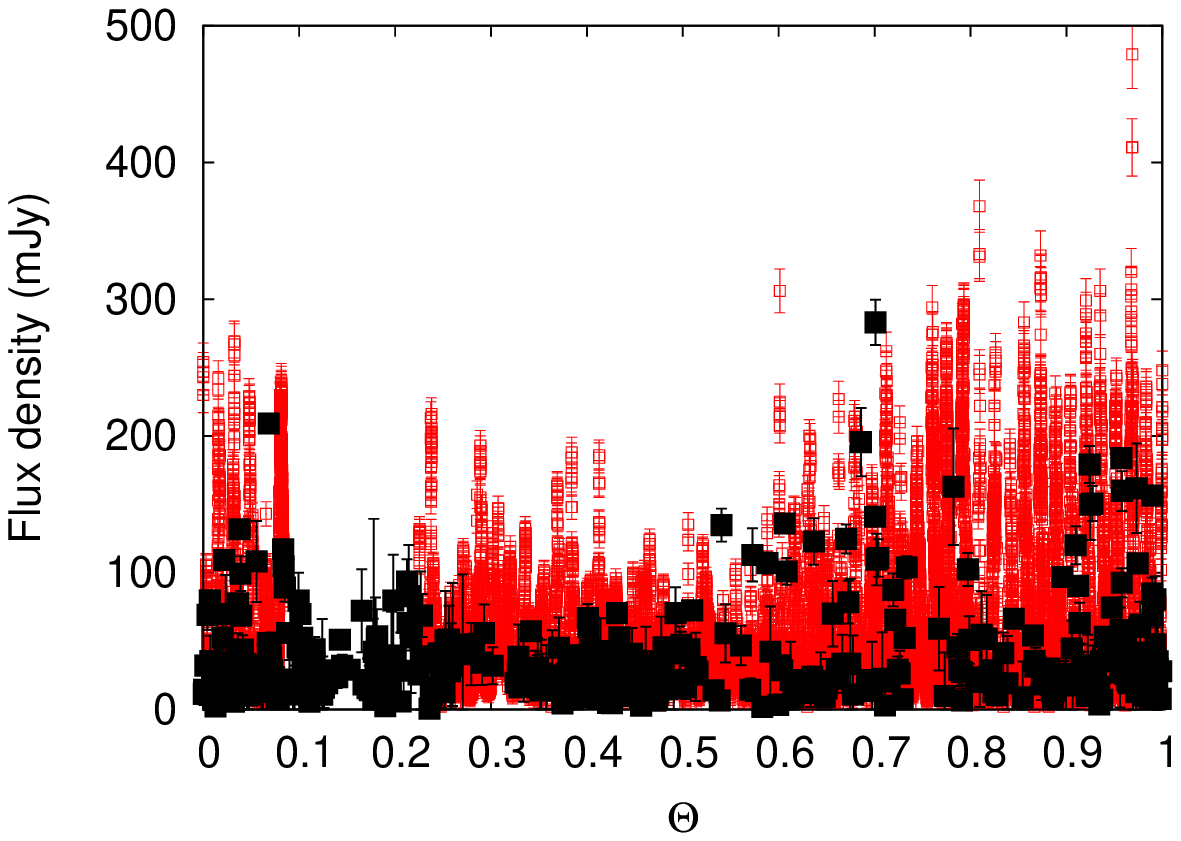}
  \includegraphics[scale=.45, clip]{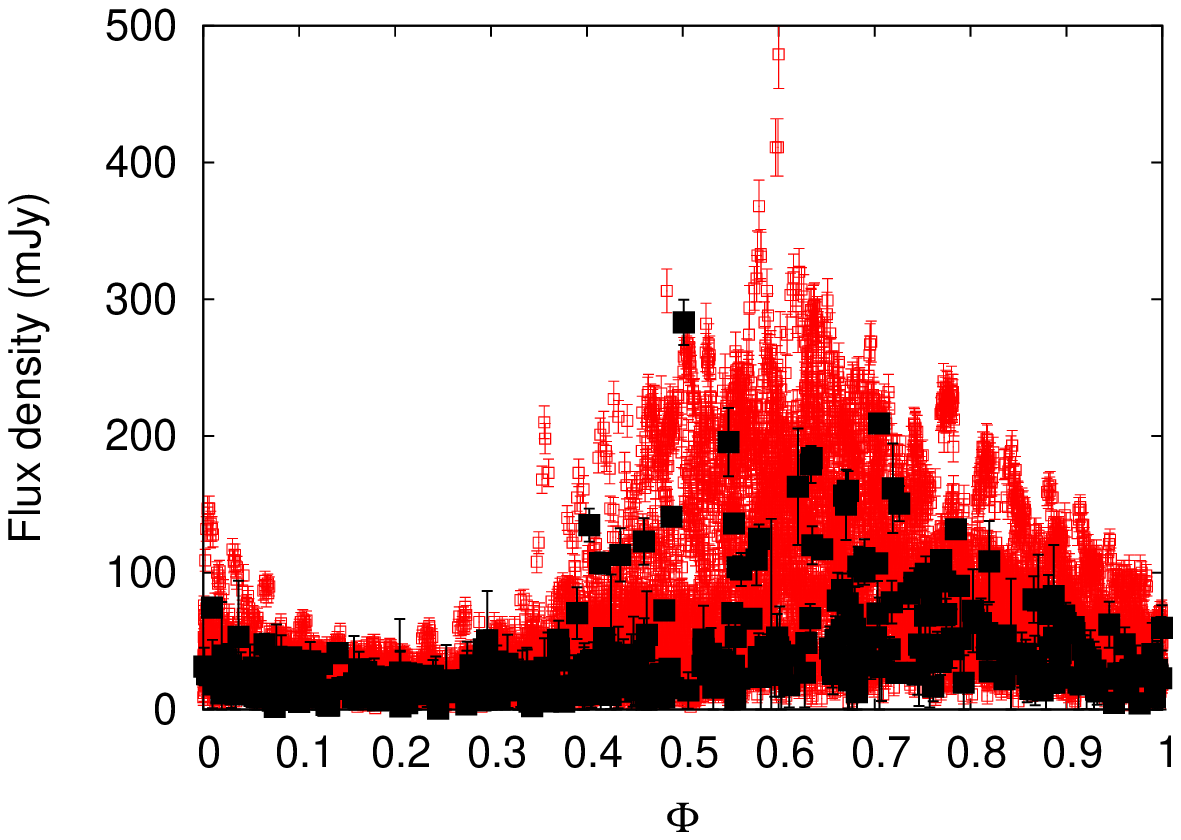}\\
  \caption{Left: OVRO data vs MJD. Centre: OVRO (filled squares,
black) data and GBI 8 GHz  data (empty squares, red) vs  $\Theta$;
          $\Theta$  is the fractional part of $(t-t_0)/1667$
          with     $t_0$=JD2443366.775  \citep{gregory02}. 
  Right:  OVRO (black) data and GBI 8 GHz data (red) vs $\Phi$, where $\Phi$ is the orbital phase,
  i.e., the fractional part of $(t-t_0)/P_1$, 
  with $P_1=26.496$ days \citep{gregory02}. 
  The largest radio outbursts 
  occur around $\Phi=0.6$. 
 At $\Phi=0.23$, i.e., at periastron,  only a low level of emission is present.}
\end{figure*}

\begin{figure}[ht]
  \centering
  \includegraphics[scale=.65, clip]{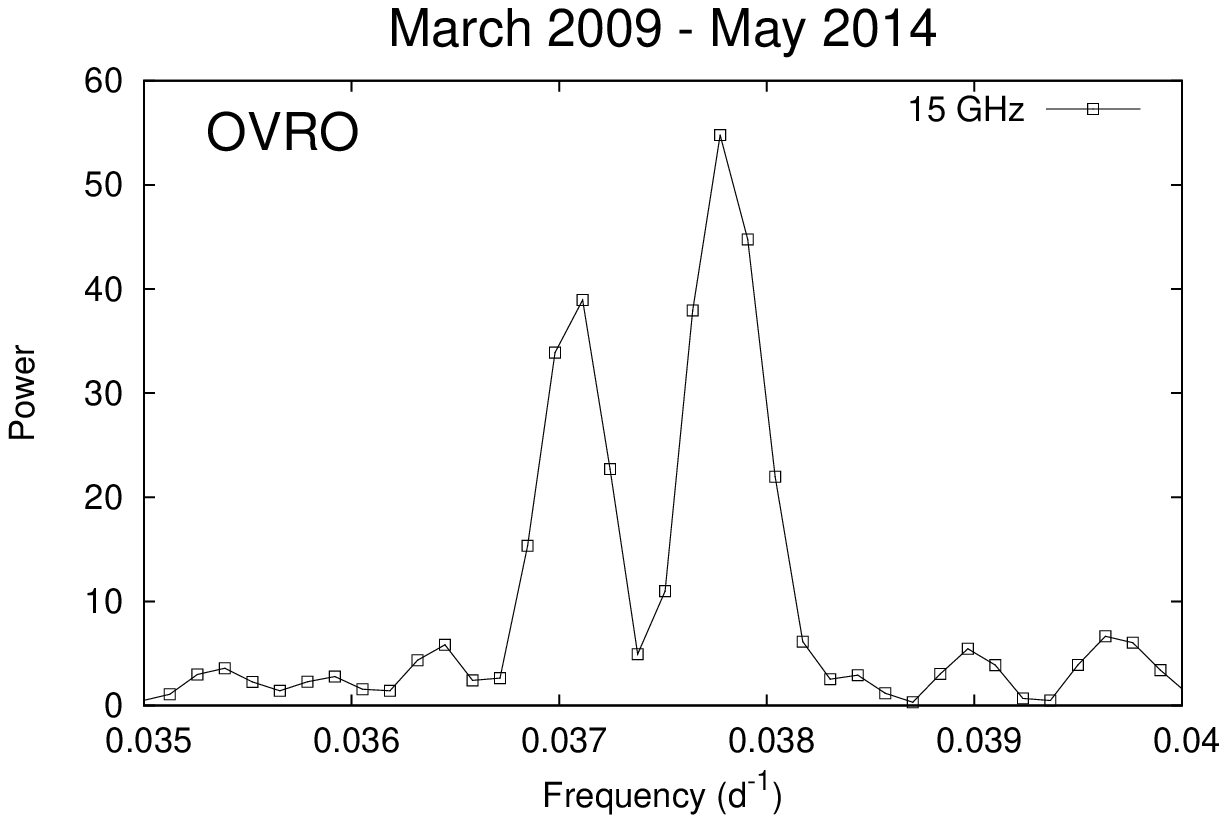}\\
  \includegraphics[scale=.65, clip]{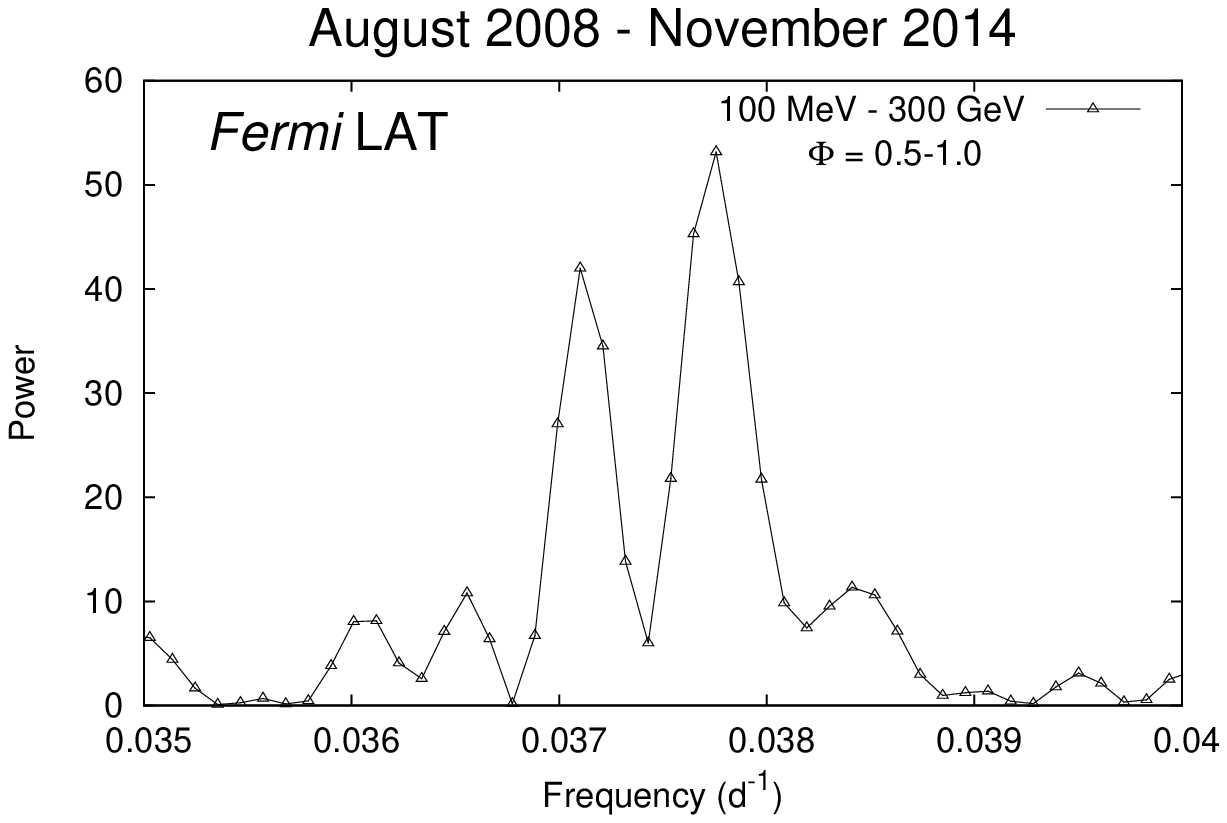}\\
  \includegraphics[scale=.65, clip]{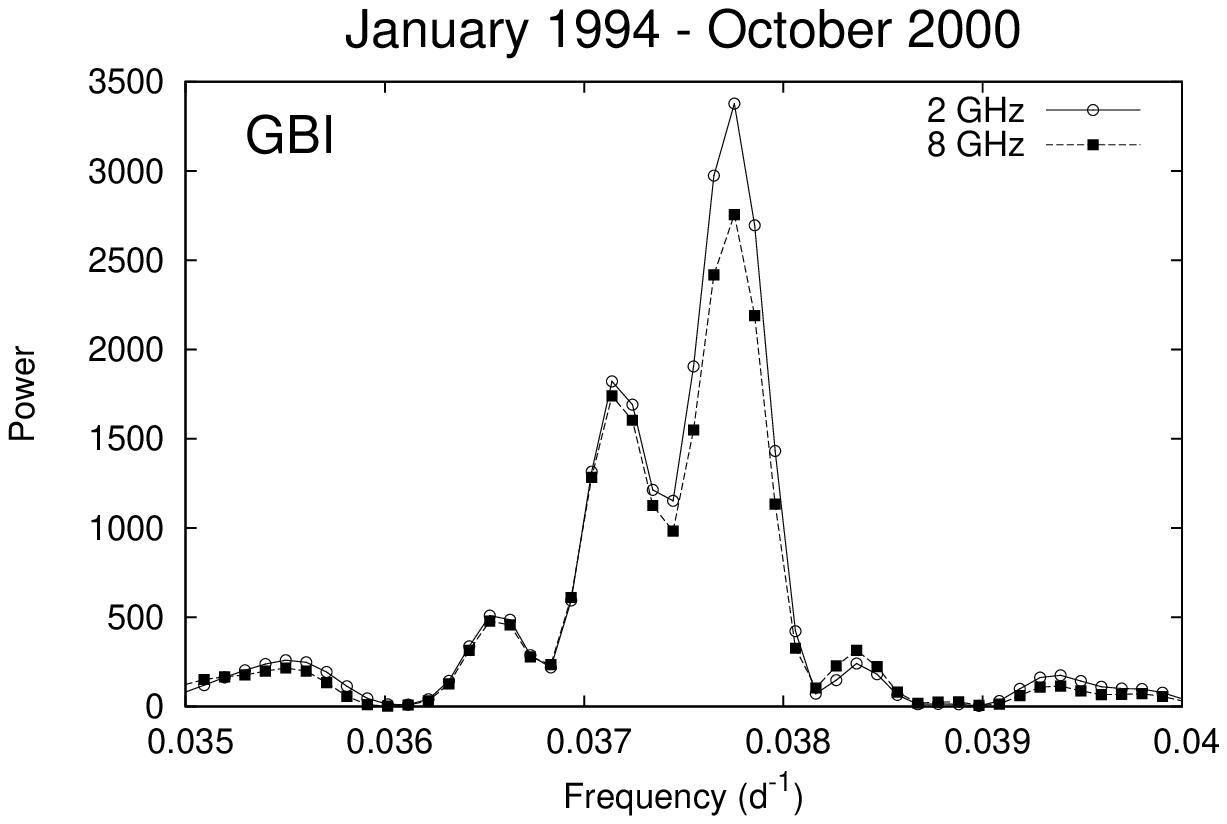}\\
  \caption{
          Lomb-Scargle periodograms of three independent long-term monitorings of
          \lsp. Top:  OVRO at 15~GHz.
          Centre: \textit{Fermi}-LAT in the energy range 100~MeV to 300~GeV
          using only data from the orbital
          phase interval $\Phi = 0.5 - 1.0$  
          \citep[see][]{jaronmassi14}. 
           Bottom: GBI at 2 GHz (empty circles) and 8 GHz (filled squares)
           \citep[see][]{massijaron13}.
          In all periodograms there are two
          periodicities with a false-alarm probability 
                  of between 0.00 and 0.01
           (see Sect. 2).
          The ratio (R) between the intensity of the two spetral features 
          at $P_1$ and $P_2$ is  different in the different periodograms and is
         R=1.9 at 2 GHz and R=1.5 at 8 GHz for GBI data, R=1.4 at 15 GHz and R=1.3 for GeV data.
  }
  \label{Fig1}
\end{figure}
Two distinct outbursts along the orbit  are well understood in the context of accretion along an
eccentric orbit  \citep{bondihoyle44}. 
The expression of the  accretion rate is
\beq
\dot{M}~={4 \pi \rho_{\rm wind}(G M_X)^2\over~v_{\rm rel}^3}
,\eneq
 where $\rho_{\rm wind}$ is the density,
  and  $v_{\rm rel}$ is the relative velocity between the compact object 
  and the Be equatorial wind.
As shown by several authors for the eccentric orbit of 
\object{LS~I~+61303,}  two peaks result:
 One peak
corresponds to the periastron passage  because of the highest density;
the second
peak occurs in the phase interval (toward  apastron)
where the reduced velocity of the compact object
compensates for the decrease in density
\citep{taylor92, martiparedes95, boschramon06, romero07}. 
 At each accretion peak,  matter is assumed to be  ejected outward in two
 jets perpendicular to the accretion disk plane as for microquasars, but for  the  two ejections 
different energetic losses for the particles occur.
Ejected relativistic electrons around periastron
suffer severe inverse Compton losses because of the proximity to the
Be star \citep{boschramon06}: the electrons upscatter stellar photons
to higher energies, and by doing so, they loose their energy and become
unable to generate synchrotron emission in the radio band. The
consequence is that around periastron a high-energy outburst caused by inverse Compton process  results, but no or negligible radio
emission \citep{boschramon06}. This first predicted gamma-ray outburst
has been confirmed by \textit{Fermi}-LAT observations \citep{abdo09,
  hadasch12, ackermann13, jaronmassi14}. 
 During the second accretion peak, the
 compact object  is much farther away from the Be star, and  inverse Compton losses
 are lower: The electrons can propagate out of the orbital
 plane and  generate synchrotron emission in the radio
 band and a minor gamma-ray outburst \citep[see the predicted outbursts in Fig. 2 of][]{boschramon06}.
Indeed, as quoted above, radio and gamma-ray  observations confirm the predicted   radio
 and  minor  gamma-ray outburst  toward apastron.
The  spectrum of the radio  outburst    has been reported to be  flat 
from 1.5 to 22 GHz by \citet[][]{gregory79}, and recently, \citet{zimmermann13} 
 observed a flat spectrum  up to 33 GHz.
These radio observations corroborate  the microquasar hypothesis  for \lsi because 
a flat radio spectrum  is indeed associated with the conical radio jet of microquasars
 \citep{fender00, fender01, kaiser06}.

In 1993, the radio-emitting source was resolved with high-resolution radio
observations showing a structure of milliarcsecond (mas) size
corresponding to a few AU at the distance of 2.0~kpc \citep{massi93}.
Successive observations revealed that the radio morphology not only changes
position angle, but it is even sometimes one-sided and at other times
two-sided
\citep{peracaula98,
          paredes98, taylor00, massi01, massi04}. 
 This suggested that \lsi might be a precessing
microquasar \citep{kaufman02, massi04}. A precession of the jet leads
to a variation in the angle between the jet and the line of sight and
therefore to variable Doppler boosting. The result is a
continuous variation in both the position angle of the radio-emitting
structure and its  flux density \citep{massi07}.
Concerning the precession time-scale, rather fast position angle
variations of almost 60${\degr}$/day measured by MERLIN observations
\citep{massi04} were  confirmed by VLBA observations 
\citep{dhawan06}, which measured a rotation  of
roughly $5\degr-7\degr$ in 2.5~hrs (i.e., $\sim
60{\degr}$/day). 
Radio astrometry resulted in a precessing period  of 27-28 days
\citep{massi12}.

A timing analysis of the GBI radio data of \lsi has revealed two rather close
frequencies: $P_1 = {1\over \nu_1} = \unit[26.49 \pm 0.07]{days}$ and
$P_2 = {1\over \nu_2} = \unit[26.92 \pm 0.07]{days}$
\citep{massijaron13}. The period\ $P_1$ agrees with the value of
$\unit[26.4960 \pm 0.0028]{days}$ \citep{gregory02} associated with the
orbital period of the binary system and with the predicted periodical accretion peak.
Period\ $P_2$ agrees well with the
estimate  by radio astrometry of 27-28~days for the precession period.

Recently,
a Lomb-Scargle analysis of \textit{Fermi}-LAT data around apastron
revealed the same  periodicities $P_{1_{\gamma}} = \unit[26.48
\pm 0.08]{d}$ and $P_{2_{\gamma}} = \unit[26.99 \pm 0.08]{d}$ 
\citep{jaronmassi14}. 
The similar behaviours of the emission  at high (GeV) and low (radio) energy
would 
imply that the emission  is caused by the same population 
 of electrons in a precessing jet.
This important result must be confirmed by simultaneous observations.
GBI observations were made from January 1994 until October 2000.
\textit{Fermi}-LAT has  observed since August 2008.
The fact that  the two archives do not even overlap  shows the persistence of $\nu_1$ and $\nu_2$
and characterizes them as  permanent features of the system \lsp.
However, to investigate the electron population, one needs simultaneous monitorings.
Fortunately, a new radio long-term
monitoring is  available. 
The Owens
Valley Radio Observatory (OVRO) 40~m telescope  monitoring at 15~GHz began
in March 2009.
In this Letter we present a timing analysis of these data. Section
2 describes  the  data reduction and the results. Section  3 presents 
our conclusions.

\section{Data analysis and results}

\subsection{OVRO observations}
Regular 15 GHz observations of \lsi{} from 54908.8 MJD (March 2009) to 56795.0 MJD (May 2014)
were carried out approximately twice per
week using the OVRO telescope 
\citep{richards11}.
The OVRO 40~m uses off-axis dual-beam optics and a cryogenic high electron mobility transistor (HEMT) low-noise amplifier with a
15.0~GHz centre frequency and 3~GHz bandwidth. The two sky beams are Dicke-switched using the
off-source beam as a reference, and the source is alternated between
the two beams in an ON-ON fashion to remove atmospheric and ground
contamination. Calibration is achieved
using a temperature-stable diode noise source to remove receiver gain
drifts, and the flux density scale is derived from observations of
3C~286 assuming the \cite{baars77} value of 3.44~Jy at
15.0~GHz. The systematic uncertainty of about 5\% in the flux density
scale is not included in the error bars.  Complete details of the
reduction and calibration procedure are found in \cite{richards11}.
OVRO data vs time, long-term, and orbital phase are shown in Fig. 1
\subsection{Timing analysis}
To
search for possible periodicities, we used the Lomb-Scargle method,
which is very efficient  for irregularly sampled data \citep{lomb76,
  scargle82}. We used the algorithms of the UK Starlink software
package, PERIOD\footnote{\url{http://www.starlink.rl.ac.uk/}}. 
The statistical significance of
a period is calculated in PERIOD following the method of Fisher
randomization as outlined in \citet{nemec85}. The advantage of using a
Monte Carlo- or randomization test is that it is distribution-free and is not constrained by any specific noise models (Poisson,
Gaussian, etc.). The fundamental assumption is that if there is no
periodic signal in the time series data, then the measured values are
independent of their observation times and could  have occurred
in any other order. One thousand randomized time series are formed, and the
periodograms are calculated. The proportion of permutations that give a
peak power higher than that of the original time series would then
provide an estimate of $p$, the probability that for a given
frequency window there is no periodic component present in the data
with this period. A derived period is defined as significant for $p <
1\%$, and a marginally significant one for $1\% < p < 10\%$ \citep{nemec85}. 
Figure 2  shows the result for OVRO data. There are two periodicities
$P_{1, \rm{OVRO}} = \unit[26.5 \pm 0.1]{d}$
and $P_{2, \rm{OVRO}} = \unit[26.9 \pm 0.1]{d}$.
The periods are both significant: they  result in the randomization test 
with a  false-alarm probability of  $p < 1\%$.
 \subsection{\textit{Fermi}-LAT}
 The timing analysis in \citet{jaronmassi14} for  the whole data set
 gave $P_2$ as  significant ($p < 1\%$) in the randomization tests
  even
 if it is a rather weak feature   in the periodograms of their
 Fig.~3 a, b, and c. 
 In other words, the timing analysis performed on the whole Fermi-data set yields that $P_2$ 
 is significant even if the spectral feature is  much
lower than that at $P_1$. 
 When the timing analysis is performed for emission in the orbital phase $\Phi$=0.5-1.0,
 as described in \citet{jaronmassi14}, the spectral feature at $P_2$ is comparable with that at $P_1$,
 but the randomization tests find  
a probability of false detection of $p=4\%$.
The $\gamma$-ray data used in that  analysis did span the time period
MJD~54683 (August 05, 2008) to MJD~56838 (June 30, 2014).
 Now we have five more months of \textit{Fermi}-LAT
 observations.
The data reduction 
was performed as reported in \citet{jaronmassi14}.
The periodogram 
for data in the orbital phase $\Phi$=0.5-1.0
shown in Fig. 2  has  the same periodicities as in  \citet{jaronmassi14},
but with the important difference that now the randomization test gives  
both frequencies as significant,  with
a  false-alarm probability of $p < 1\%$.
\subsection{385 days of GBI data vs  RATAN results}
\begin{figure}[h]
  \includegraphics[scale=.75, clip]{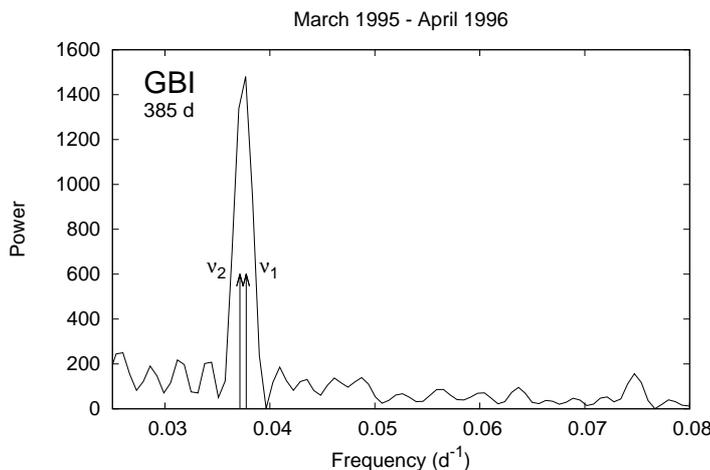}\\
  \caption{
     Lomb-Scargle periodograms of 385 days of GBI radio data at 2~GHz to be compared with
     the periodogram in \citet{trushkin14} also of 385 days.
     The spectral resolution of only 0.00065 d$^{-1}$ cannot resolve the 
     separation of  0.0006 d$^{-1}$ between 
     $\nu_{1} = \unit[0.03775 \pm
         0.00010]{d^{-1}}$ and 
         $\nu_{2} = \unit[0.03715 \pm  0.00010]{d^{-1}}$ determined by \citet{massijaron13} with 6.7 yr of GBI data.
   }
  \label{fig:Box1SC}
\end{figure}

The 
frequency resolution in a periodogram
is related to the inverse of the overall time interval of observations.
The resolution for GBI of $\Delta \nu =
\unit[0.0001]{d^{-1}}$  covers six times the difference in frequency
between $\nu_1=\unit[0.03775 \pm 0.00010]{d^{-1}}$ and $\nu_{2} = \unit[0.03715 \pm  0.00010]{d^{-1}}$ 
determined by \citet{massijaron13}.  Indeed, the two spectral features
are evident in the GBI periodogram shown here  in Fig. 2.
Very recently,  \citet{trushkin14} reported that the  timing analysis of RATAN-600  radio telescope 
data  did not  find the two frequencies.
The reported daily monitoring with the
RATAN-600 radio telescope includes the time from 16 November 2013 (MJD 56612) to 6
December 2014 (MJD 56997), that is, 385~days. This  corresponds to a
frequency resolution of 0.00065~d$^{-1}$ that is even higher than the
difference in frequency between $\nu_1$ and $\nu_2$ of 0.00060~d$^{-1}$. 
We tested the 
insufficient frequency
resolution by performing  the timing analysis
on  385 days of  GBI data. 
The resulting periodogram,
shown in Fig. ~\ref{fig:Box1SC}, is well comparable with that of  \citet{trushkin14}.
The two frequencies $\nu_1$ and  $\nu_2$ are plotted in Fig. 3 by arrows.
The low spectral resolution is still unable to resolve the frequency separation.

\begin{table*}
        \center
        \begin{tabular}{|lcccccc|}
                \hline
                $P_1 \pm \Delta P_1$&$P_2 \pm \Delta P_2$& Ratio&$P_{\rm beat} \pm \Delta P_{\rm beat}$&Monitorings&Duration&Reference\\
                                days&days& &days&&years&\\
                \hline
                \hline
                  $26.5 \pm 0.1$&$26.9 \pm 0.1$& 1.4& $1782 \pm 630$& OVRO (15 GHz)&5.2& this work\\
                \hline
                $26.49 \pm 0.07$&$26.92 \pm 0.07$& 1.9& $1658 \pm 382$& GBI (2 GHz)&6.7& a\\
                $26.49 \pm 0.07$&$26.92 \pm 0.07$& 1.5& $1658 \pm 382$& GBI (8 GHz)&6.7& a\\
                $26.48 \pm 0.08$&$26.99 \pm 0.08$& 1.3& $1401 \pm 311$& Fermi-LAT (0.1 - 300 GeV), $\Phi =0.5 - 1.0$&6.3& b\\
                \hline
        \end{tabular}
        \caption{Timing analysis results. The ratio between the intensity of the two spectral features at $P_1$ and $P_2$ in the periodograms of Fig. 2 is given in the third column.  $P_{\rm beat}$ is defined as  $P_{\rm beat}= (P_1^{-1}-P_2^{-1})^{-1}$.
        Column six reports the duration of the related monitoring. The references are a: \citet{massijaron13} and b:  \citet{jaronmassi14}.}
\end{table*}

\section{Conclusions and discussion}

The spectral resolution in the timing analysis of OVRO data is able to distinguish the two spectral features
found in GBI and Fermi-LAT data (Table 1). 

These two periodicities are physically related to the periodical ejection ($P_1$) of electrons in a particular orbital phase
toward apastron and to the precession ($P_2$) of the jet. The beating of these two periodicities  probably causes
the so-called long-term modulation.  
The peak flux density of the periodical radio outburst
exhibits  in fact a   modulation of 1667$\pm$8~d \citep{gregory02}.
  \citet{massijaron13} have shown the consistency between  the long-term modulation and 
  the beating of the two frequencies determined 
  in the GBI timing analysis, that is
, $P_{\rm beat}= (P_1^{-1}-P_2^{-1})^{-1}$ (Table 1).
  Moreover, \citet{massitorricelli14} have shown that a  physical
  model for \lsi{} of synchrotron emission from a precessing ($P_2$)
 jet, periodically ($P_1$) refilled with relativistic particles, produces 
a maximum  when 
the jet electron density is at
its maximum and the approaching jet forms the smallest
  possible angle with the line of sight. This coincidence of the
    highest number of emitting particles and the strongest Doppler
    boosting of their emission occurs with  a period of $(P_1^{-1}-P_2^{-1})^{-1}$, 
    that is, $P_{\rm beat}$ , or the long-term periodicity.

The open question now is: why is the precession so close to the orbital period?
While the microquasar SS433 and several  known precessing X-ray binaries
\citep{larwood98} have a precession period longer by an order of magnitude than the orbital period (as predicted for  a tidally forced precession by the companion star),  
the different case of the microquasar  GRO J1655$-$40 was discovered
 in   1995  \citep{hjellmingrupen95}.
 This system has
 an orbital period of   2.601$\pm 0.027$~days \citep{bailyn95}.
 \citet{hjellmingrupen95}  discovered  a radio jet in this object with a  precessional period  of
 3.0$\pm$0.2 days  \citep{hjellmingrupen95}.
\lsi seems to be the second case of this new class of objects with orbital and precession periods that are rather close to each other.

\begin{acknowledgements}

The OVRO 40 M Telescope Monitoring Program is supported
by NASA under awards NNX08AW31G and NNX11A043G, and by the NSF
under awards AST-0808050 and AST-1109911.
We would like to thank J\"urgen Neidh\"ofer, Giammarco Quaglia, and Eduardo Ros  for helpful comments.
This work
  has made use of public \textit{Fermi} data obtained from the High
    Energy Astrophysics Science Archive Research Center (HEASARC),
      provided by NASA Goddard Space Flight Center.
The Green Bank Interferometer is a facility of the National
Science Foundation operated by the NRAO in support of NASA High Energy
Astrophysics programs. 
\end{acknowledgements}

\bibliographystyle{aa}

\end{document}